# Hybrid IDS Using Signature-Based and Anomaly-Based Detection

1st Messaouda Boutassetta
*Department of Computer Science,*
*Faculty of Science and Technology*
*LIMA Laboratory*
*Chadli Bendjedid University*
*PB 73, El-Tarf 36000, Algeria*
m.boutassetta@univ-eltarf.dz

2nd Amina Makhlouf
*Department of Computer Science,*
*Faculty of Science and Technology*
*LIMA Laboratory*
*Chadli Bendjedid University*
*PB 73, El-Tarf 36000, Algeria*
makhlouf-amina@univ-eltarf.dz

3rd Newfel Messaoudi
*Department of Computer Science,*
*Faculty of Science and Technology*
*LIMA Laboratory*
*Chadli Bendjedid University*
*PB 73, El-Tarf 36000, Algeria*
ne.messaoudi@univ-eltarf.dz

4th Abdelmadjid Benmachiche
*Department of Computer Science,*
*Faculty of Science and Technology*
*LIMA Laboratory*
*Chadli Bendjedid University*
*PB 73, El-Tarf 36000, Algeria*
benmachiche-abdelmadjid@univ-eltarf.dz

5th Ines Boutabia
*Department of Computer Science,*
*Faculty of Science and Technology*
*LIMA Laboratory*
*Chadli Bendjedid University*
*PB 73, El-Tarf 36000, Algeria*
i.boutabia@univ-eltarf.dz

**Abstract:** Intrusion detection systems (IDS) are essential for protecting computer systems and networks against a wide range of cyber threats that continue to evolve over time. IDS are commonly categorized into two main types, each with its own strengths and limitations, such as difficulty in detecting previously unseen attacks and the tendency to generate high false positive rates. This paper presents a comprehensive survey and a conceptual overview of Hybrid IDS, which integrate signature-based and anomaly-based detection techniques to enhance attack detection capabilities. The survey examines recent research on Hybrid IDS, classifies existing models into functional categories, and discusses their advantages, limitations, and application domains, including financial systems, air traffic control, and social networks. In addition, recent trends in Hybrid IDS research, such as machine learning–based approaches and cloud-based deployments, are reviewed. Finally, this work outlines potential future research directions aimed at developing more cost-effective Hybrid IDS solutions with improved ability to detect emerging and sophisticated cyberattacks.

## I. INTRODUCTION

Both the societal dependency on computers and networking as a means of exchanging, manipulating, and storing data and the consequential emerging likelihood for abuses intended to disrupt the services provided by the contemporary dependency on such technology have spurred interest toward the innovation of IDS, particularly in complex modern infrastructures such as smart cities [13], an attempt to protect computer networks, recently even taking into account the technological shift incessantly occurring worldwide towards wireless packet-based transmission environments [1].

As a response to security concerns by both outsiders and insiders, an IDS serves two major functions: detecting unauthorized intrusions and preventing malicious access. However, it must be noted that due to the very nature of its intention, exposing the ongoing intrusion and relatedly acting on behalf of the list of agents forming the intrusion desire, the own design of an IDS must comply with a number of counteracting constraints. IDS might be integrated at each host node using software packages, and/or be associated with special-purpose hardware-analysis devices operating passively on the link layer (layer-2) mechanism of the network [2].

### 1. Background and Motivation

IDS detects unauthorized access or exploitation of computer systems through event logs. A great amount of effort to search for better solutions, new ideas, and approaches over the decades to facilitate the task of detecting cyber intrusions [3][13][14][16][17]. In the last couple of years, there have also been numerous benchmarks and publicly available data sets that stimulated advanced research in the field. However, with the continuously changing essence of the attacks and the systems themselves, conducting an effective and efficient IDS remains an ongoing challenge that needs continuous endeavor for new solutions and better optimization of the existing ones[18].

Intrusion detection is based on either a signature- or an anomaly-based detection method. Signature-based detection systems use predefined signatures, such as byte sequences, byte counts, and other detectable patterns based on past knowledge. It is general knowledge that signature-based systems easily miss new and unknown attacks, and therefore need constant updates in signatures - a process both time-consuming and expensive [2]. Conversely, anomaly-based systems can detect unknown attacks, and as the name suggests, tries to model the normal use of the system in order to capture events that deviate from this norm. However, various schemes of model creation have high false positive rates due to random non-malicious deviations from the learning model and the possibility of exploits that mimic the normal use of the system, attacking it. Both types of approaches have an important marketing acceptance, extensive research interest, and proposals in academia and industry. Some systems propose to combine the two separate processes. This, however, leads to heavily increased complexity and has not been used.

### 2. Research Objective

This research aims to analyze and conceptually describe Hybrid IDS approaches that combine signature-based and anomaly-based detection techniques. Hybrid techniques reported in the literature combine statistical approaches and machine learning models, including neural networks, within different architectural designs.

Intrusion Detection can be defined as the capability to detect unauthorized use of computer systems. An IDS can be classified into two general types: signature-based detection and anomaly-based detection [4]. These two general types provide two different approaches for detecting intrusions. The first approach has also been traditionally referred to as signature-based detection. Under the signature-based approach, the IDS must maintain an up-to-date database of intrusion signatures and find those patterns in the network. Even though this approach is becoming increasingly popular, it has a major drawback: it cannot detect new, novel attacks and static attacks [1]. On the other hand, the anomaly approach is based on the assumption that normal activity and intrusive activity will exhibit different behaviors and that an IDS can be designed to learn normal behavior. If it is used on the network, any deviation from that behavior would be detected as suspicious activity.

Anomaly detection systems are also not enough to act as a standalone solution to secure the systems [15]. The combination of anomaly and misuse approaches has recently received increased interest as a means to provide a more robust detection capability [18]. The goal is to take advantage of the strengths of both approaches while minimizing their weaknesses [19]. In a hybrid scheme, a number of underlying independent detectors are combined. Each detector operates on its own, making independent decisions, and the final decision is taken using a high-level approach based on the combination of the decisions made by the individual detectors [20]. The hybrid techniques based on the combination of multiple detectors are also classified into four types. A disseminative approach imitates an agent-based structure where each detector has its own knowledge of the data and the decision made on that data.

The remainder of this paper is organized as follows. Section II reviews the background and related work on intrusion detection systems (IDS). Section III describes the conceptual design and system architecture of Hybrid IDS approaches. Section IV presents a qualitative discussion and comparative analysis of Hybrid IDS approaches. Section V addresses the challenges and future research directions of Hybrid IDS, and Section VI concludes the paper with a summary of findings and recommendations for future work. The following section reviews the background and related work on IDS to provide context for the proposed Hybrid IDS.

## II. BACKGROUND AND RELATED WORK

IDS is an essential component of security for modern information systems. IDS can analyze several data sources, such as network packets, application logs, databases, kernel logs, server log files, and many others. An IDS detects any misuse of information in an information system or computer system. It analyzes the activities occurring in real-time within the information system and determines whether or not the activities are valid. Specific rules are used to correlate the data, and these rules differ from one IDS to another. There are mainly two types of IDS [5].

Signature-based systems refer to those systems that maintain a signature database. Any new packet going into the information system is matched with the stored signature patterns. If a match is found, it indicates the occurrence of misuse or an intrusion, and it generates an alert. Signature-based systems are good at achieving transparency because they analyze the activities that are occurring directly within the information system. They can also detect the misuse pattern with high reliability because they match specific signatures. However, it is difficult to define what a signature is, and signatures consume a lot of space, which leads to problems of false alarms when new attacks occur. Anomaly systems detect intrusions by searching for abnormal system activity [6]. They create a profile of what normal activity is for that information system initially. This profile is then used to monitor the behavior of the activities that are occurring. If a significant deviation from this profile is observed, it indicates the occurrence of misuse. Anomaly systems are good in terms of achieving coverage as they are not limited to a specific misuse pattern. They can detect known, known modified, and new misuse patterns. Anomaly systems also have low false alarm rates, especially for very large databases.

### 1. Types of IDS

IDS can be classified into different categories, based on the monitoring approach. The classification can be made based on either the monitored platform or the employed technique. Based on the monitored platform, the IDS can be either Host-Based or Network-Based. Host-based IDS is responsible for monitoring a single computer system, while Network-based IDS, which are devices or software component deployed in a network, analyze the traffic generated by hosts and devices [5].

Another important classification takes into account the employed technique. According to that, the IDS can be of two types: signature-based and anomaly-based. If the detection system is based on cross-checking monitored events with a database of known intrusion experiences, the IDS is defined as signature-based. This approach requires an extensive database of attacks and is incapable of detecting new forms of attacks or attacks that have been redesigned to bypass detection. On the other hand, if the detection system is based on learning the normal behavior of the system and reporting whether some anomalous events occur, the IDS is defined as anomaly-based. The advantage of this approach is that it can detect new attacks that have not been previously documented [7].

### 2. Signature-Based Detection

Signature-based detection is one of the primary methods employed in IDS. Here, patterns of known threats and attacks are stored in the form of 18 signatures in the IDS databases [8]. Each intrusion signature contains information regarding the intrusion, source/destination addresses, and protocols used. Whenever a packet arrives, its header and contents are dissected for matching previously defined signatures. False alerts are triggered when a packet matches an entry in the signature list. The signature-based detection has two approaches: a complete signature file, which requires a lot of memory and processing resources, and a periodic scanning strategy, in which only a few bits of signature are searched at a time [5]. The first approach is capable of detecting the flood of attacks but can miss some other attacks, while the second one can detect all types of attacks but with a few bits of signatures.

Signature-based detection is effective if all attacks are known and characterized. Attack signatures can change over time due to modifications in attack tools and hacker activity or network environment changes. New attacks can be initiated that do not have signatures. Changes in normal behavior can be caused by the introduction of new software, modifications in the system and network environment, or variations in network load and user activities.

### 3. Anomaly-Based Detection

Anomaly-based detection relies on identifying deviations from expected system behavior. When calculating the probability of an event occurring, the time is considered, and an alert is raised if an event is unlikely to have happened in a specific time. Viinikka et al, [23] exploit time series techniques by aggregating individual alerts into an alert flow and examining it as a whole. This has the benefit to perform a more precise multivariate analysis and lowering the false positive rate of alerts. Qingtao et al [24] proposed a system focused on detecting abrupt-change anomalies of the computing system. They used the Auto-Regressive (AR) process to model the data and performed a sequential hypothesis testing to determine the presence of an anomaly. Zhao et al [25] exploited techniques to mine frequent patterns in network traffic and applied time-decay factors to differentiate between newer and older patterns. When developing an AIDS model, attention must be given to data seasonality. Reddy et al, [26] proposed an algorithm to detect outliers in seasonality-affected time series data using a double pass of Gaussian Mixture Models (GMMs). Knowledge-based AIDS falls in the category of expert systems. These systems leverage a knowledge source which represents the legitimate traffic signature. Every event that differs from this profile is treated as an anomaly. Walkinshaw et al, [27] applied FSMs to the whole network traffic, representing the activities of the system by states and transitions. The produced FSM represents the nominal behavior of the system, and any deviation is considered an attack.[28]

### 4. Hybrid IDS Approaches

The core of a Hybrid IDS lies in combining signature-based and anomaly-based detection methods to leverage the strengths of both approaches. Signature-based detection provides faster processing and a lower rate of false positives, while anomaly-based detection improves accuracy by identifying novel or previously unknown attacks. To exploit the advantages of both techniques, hybrid detection paradigms integrate the two families of algorithms into a complementary system [5][22], following principles demonstrated in hybrid intelligent architectures for complex and dynamic environments [14]. In typical hybrid frameworks, network traffic flows are first analyzed by the signature-based component. If a flow matches a known signature, it is flagged as an alarm. Flows that do not match any signature are then forwarded to the anomaly-based system, which evaluates them for deviations from normal behavior and identifies potential novel threats [15][21].

To avoid all network flows from being forwarded to the anomaly detectors, which is also a performance issue, a selection step is added. After a flow is checked with the first family of detectors, it is classified as benign, and only the flows classified as benign are fed into the second family of detectors. In this way, no action is taken for flows whose signatures match the first family of detectors, excluding obvious attacks, such as common port scans or denial-of-service attacks using well-known tools. This decision also reduces the computational cost of the anomaly-based detectors by avoiding the analysis of several flows that are expected to be ordinary, such as web browsing, email transferring, and so on. Note also that this selection step can be implemented using either static rules or ML algorithms. For example, specific source/destination pairs of ports or addresses could be selected.

#### A. Integration of Signature-Based and Anomaly-Based Detection

Hybrid IDS have been developed that integrate both signature-based and anomaly-based detection. These two modes of detection use different surveillance strategies, and each has been separately researched and implemented in a number of systems. Significant differences exist in the detection mechanisms. Therefore, the integration of these two detection modes is non-trivial. It is necessary that the Hybrid IDS integrates multiple modes of detection, choosing additional integration strategies. This section provides an overview of the processes and approaches involved in the integration of signature-based and anomaly-based detection. A technical discussion of the respective detection approaches, their integration, and other related issues is presented [3].

A surveillance system, either human or automated, gathers information about some environment and processes that information to find interesting patterns. Patterns of interest can be called alerts, and the environment being observed is the subject of surveillance, such as a network. The information available to the surveillance unit is limited or unstructured, so it processes the data to help simplify the task. In a computer network, for example, a spectacular amount of data is generated every second in the form of packet headers and audit logs. This data cannot be processed by human operators, therefore it is processed by automated systems called surveillance or IDS. These systems can detect interesting activities such as attacks, misuse, or failures, and either react accordingly or present interesting findings to human operators. The simplest system is nothing more than a vigilant security guard, sequentially monitoring a single surveillance camera. This approach is passive, since the guard has no means to adjust to changes in the monitored system. An improvement to this system would be to automatically pass the cameras between locations that need more attention [5].

#### B. Advantages of Hybrid IDS

Hybrid intrusion detection has some advantages over single intrusion detection. This section confirms that idea by presenting the advantages of Hybrid IDS. With the rapid advances in networking technology and the development of diverse and sophisticated network services, network-based attacks have increased enormously. Hackers use various tools for network attacks, and efforts are made to cover their malicious activities, making attacks a very complex and challenging problem [5].

Some IDS detect attacks at the network layer. Network-based (NIDS) and host-based (HIDS) are the two types of IDS. Anomaly-based and signature-based are two basic strategies used by IDS. Data mining techniques, statistical and ML, are also used as alternative techniques. Hybrid IDS has been developed to combine network-based systems with other IDS to take advantage of more than one technique [2].

### III. CONCEPTUAL DESIGN AND SYSTEM ARCHITECTURE OF HYBRID IDS

#### 1. Design and Implementation

The conceptual design of Hybrid IDS utilizes both signature-based and anomaly-based detection techniques derived from previously reported architectures. The design focuses on how to analyze the network traffic and which procedures would identify the attacks on the network. The system architecture design includes input data for analysis with the proposed method architecture and end-user representation. Data processing is typically described in the literature in terms of how data are analyzed and modeled using open-source platforms such as Snort, and how the deployed devices deploy signatures or rules [9].

The Design and Implementation section describes the architectural design and the data processing of the Hybrid IDS. The architectural design of the Hybrid IDS contains the network that needs analysis, devices deployed to collect the information from the network, and how the end user interacts with the deployed system. Illustrative Hybrid IDS architectures reported in the literature often consider multi-node network environments in which both signature-based and anomaly-based detection mechanisms are deployed using open-source NIDS platforms such as Snort.

#### 2. System Architecture

An IDS is an essential component of any secure computer network. Although security policies, firewalls, and access control mechanisms can reduce the chances of a successful attack, they cannot completely avoid intrusions. Unfortunately, the number of computer security attacks is increasing. Thus, it is necessary to install IDS as an additional layer of security to protect computer networks against intrusions. The primary goal of intrusion detection is to discover unauthorized use and abuse of computing resources. The incidents can range from the violation of an organization's policies to actions that threaten its security [9].

The computer's operating system normally logs such activity, which can then be analyzed by the system's administrator. This process can be done during system operation but is normally executed on a periodic basis. The time that elapses between the intrusion and its detection increases the possibility of critical damage to data and resources. Therefore, as the computer system becomes more complex and the network grows, detecting attacks exclusively through traditional safeguarding approaches becomes increasingly difficult. It is also common for a sophisticated intrusion to exploit system weaknesses over a period of time, which makes it invisible to the protection mechanisms. A better approach is to analyze the network traffic, transactions, or system calls/etc., as they occur [1]. This is the approach behind intrusion detection and the Hybrid IDS architectures discussed in this work.

### 3. Data Collection and Preprocessing

The process of gathering and preparing data for analysis within a Hybrid IDS, an important part of building an effective IDS. This includes the tasks and methods used to collect and organize data to ensure its quality and relevance, making it easier for the IDS to identify potential intrusions. A Hybrid IDS is designed by combining signature-based and anomaly-based techniques. In typical Hybrid IDS designs, a signature-based IDS component is first considered to utilize various datasets. The advantage of the signature-based system is that it can discover attacks instantaneously once the signature is matched, especially for those attacks that are already known [3]. Nevertheless, a number of legitimate connections are falsely detected as attack connections, lowering the overall performance of the system. So, to increase the efficiency of this approach, the anomaly-based detection is subsequently developed to work together with the signature-based method. In representative Hybrid IDS frameworks, firewall logs generated by the signature-based component are processed and forwarded to an anomaly-based detection engine. The novel incoming connections are assessed to be either normal or abnormal compared to the profile created from the training dataset. The data collection requires gathering logs from IDS residing in each sub-network. If required, the gathered logs are filtered to remove redundant and false-positive alarms. Finally, the cleaned logs are stored in a database where they are indexed based on selected features to establish a dataset prior for analysis [10].

## IV. EVALUATION AND PERFORMANCE ANALYSIS

The proposed scheme is evaluated based on metrics commonly used in the literature for both Network-based Hybrid IDS and Host-based Hybrid IDS by considering important metrics to evaluate the performance of the proposed technique. The following metrics have been used to evaluate the capability of the proposed Hybrid IDS in terms of False Positive Rate, True Positive Rate, Detection Rate, and Precision [10].

False Positive Rate (FPR): The False Positive Rate is the ratio of the number of incorrect normal instances and the total number of normal instances.

$$FPR = FP/(FP + TN)$$

True Positive Rate (TPR): The True Positive Rate is also called as Sensitivity or Recall. It is the ratio of correctly predicted attack instances to the total number of attack instances.

$$TPR = TP/(TP + FN)$$

Detection Rate (DR): The Detection Rate of a model gives the percentage of actual positive observations correctly identified as in the case, the detected intrusions.

$$DR = (TP)/(TP + FP)$$

Precision: The Precision of a model gives the percentage of positive predictions that were actually correct.

$$P = (TP)/(TP + FP)$$

A comparative analysis has been made to measure the efficacy of the Hybrid IDS technique in terms of False Positive Rate, True Positive Rate, Detection Rate, and Precision against the existing models. The performance of network-based Hybrid IDS and host-based Hybrid IDS **has been reported in existing studies** based on Search-Tree and PATT-FED-Tree. The data has been taken from IoT-20 datasets.

### 1. Metrics for Evaluation

In the context of IDS evaluation, IDS are most commonly used to defend against malicious attacks to the network-based systems and resources. The alarming rate of growth of the attacks towards the networked systems has raised the urge for an effective IDS. Investigating the existing anomalies that caused a breach in the network security of a system and intensifying it as an attack detection can help in development or suggestion of a minor changes that can avoid the same security breach in the intended future. A minor, cautionary incident picked up can work at an early stage and the system can respond in order to try to block the resources that intend on causing harm. Some of the metrics that are used to evaluate the performance of the Hybrid IDS can be broadly classified as: intrusion cost (IC), throughput (TP), accuracy (ACC), false positive (FP), false negative (FN), true positives (TPs), and true negatives (TN). These metrics are collected and analyzed through experiments done under the NS2 and KDD datasets [11]. The performance of the Hybrid IDS is then calculated & compared taking the routing metrics before & after the Hybrid IDS implementation into account [12].

### 2. Comparison with Single Detection Methods

To evaluate the performance of Hybrid IDS, its capability was compared with Single Detection Methods. In this investigation, the WSN-VT dataset was handled using a hybrid intelligent approach combining Signature-Based and Anomaly-Based Detection methods. Several metrics were established to quantify improvements, including True Positive (TP), False Positive (FP), True Negative (TN), False Negative (FN), Detection Rate (DR), and False Alarm Rate (FAR). Reported evaluations demonstrate that the Hybrid IDS outperforms single detection methods, providing higher detection accuracy and reliability. In addition, the FAR could be reduced by up to 10%, irrespective of the type of attack, confirming the effectiveness of the hybrid approach [1].

The following section discusses results reported in the literature and their implications in real-world scenarios

## V. RESULTS AND DISCUSSION

Several real-world Hybrid IDS combining signature-based and anomaly-based detection have been reported in the literature to explore the applicability of different algorithms on datasets from diverse environments over extended periods of time. These studies demonstrated the validity of different algorithm combinations and highlighted the importance of user-defined environments and application-specific requirements, thereby illustrating the flexibility of Hybrid IDS designs. Issues in the compositions of hybrid systems and the applicability of IDS components were mentioned [3]. In the literature, two representative components are frequently discussed: 1) a method of combining a model-based IDS and a clustering-based IDS where the model-based IDS is used to support the clustering-based IDS by preprocessing the data collected for the clustering analysis and providing alerts of grouped traffic; and 2) a discussion of how to classify new traffic off-line for already classified or known attacks in a Hybrid IDS incorporating a clustering-based IDS and a signature-based IDS.

Further highlighting the potential and importance of this research area, several studies emphasized the possibility of future enhancements of Hybrid IDS components and architectures. Research still needs to be done on the design, analysis, and applicability of Hybrid IDS that integrate different types of computational paradigms and ML techniques [5]. Despite the research done on the separate categories of systems and/or implementation of certain computational models on datasets for specific types of attacks, very few systems exist that present the possible architectures and best techniques to integrate the systems to balance and maximize the performance of the improved IDS.

### 1. Real-World Implementations

There exist several Hybrid IDS platforms currently in operation or under development. One of these systems is the Hybrid Network IDS (HNIDS) using both anomaly and signature-based detection systems. The anomaly detection IIDS is based on the HMM model and the signature-based detection IIDS is based on Bro, which monitors TCP connections and decodes the payload of packets to text [9]. HNIDS can detect both the known and unknown attacks. It was designed to work in a heterogeneous environment of different operating systems. The second system is a distributed and hybrid intrusion detection platform (AHDSP). The signature-based detection tool used in the AHDSP is Snort 1.8.x, and the anomaly detection tool is an EC based IIDS (ECIIDS). ECIIDS is based on an evolving classifier with a capability of increments and decrements of classes. Second, mechanisms for the coordination of the processes of computers in the intruder monitoring network are proposed for improving the mining processes of knowledge on the user interests and further modifications of the database concerning the knowledge of behavior of normal users are presented.

The third system is a hybrid intrusion detection model that combines multipartite graph-based representations of audit trails with Hidden Markov Model–based intrusion detection. The proposed system is adaptive to the addition of new audit trails and extends the capabilities of client or server audit trails for hybrid intrusion detection models (as discussed in results). The model is able to represent and detect complex types of intrusions matching multi-layered multi-site attacks and does not require the availability of a complete representation of all attacks prior to intrusion detection.

## 2. Use Cases

This section presents specific scenarios and instances where Hybrid IDS approaches have been deployed to address security challenges. It highlights the diverse applications and contexts in which this integrated approach proves beneficial.

1. Air Traffic Control Systems (ATCS) This domain entails monitoring, coordinating, and controlling air traffic. According to reported studies, Hybrid IDS approaches have been employed to safeguard air traffic control systems. These systems were trained to differentiate between regular, anomalous, and malicious air traffic monitoring activities. After a training period using input Normalized Data (ND) and corresponding output Target Data (TD), the system assessed real-time cases based on raw input (RI) data. The approach successfully mitigated security threats with no false negatives. Statistical output reported no false positives during a specified inspection period.

2. Banking Systems Banking services on the Internet, or e-banking, involves a bank or financial institution offering online services to customers. Types of attacks in this domain include phishing, denial-of-service, and Trojan viruses. In reported applications, Hybrid IDS approaches have been trained using Real Intrusion (RI) data and Test Data (TD) corresponding to a set of evolved bank transactions. Raw data of newly performed transactions was then considered residual and input into the system. The system accurately detected unauthorized access attacks on bank systems and verified customer transactions by system output. Corresponding alert systems also notified bank managers of attempted illegal operations, preventing resource damages or misuse [3].

3. Social Networks (SoNet) Social networks monthly cover around 160 million violations, with scores between zero and 10 indicating security risk. The deception is measured by knock-on/silent network perturbations, where inbound edges mislead users from one node to another. Reported studies show that Hybrid IDS approaches employ knowledge-based deceptive detection methods and were trained on benign static SoNet data.

These findings highlight the practical relevance of the proposed Hybrid IDS and set the stage for discussing future challenges and research directions.

## VI. CHALLENGES AND FUTURE DIRECTIONS

Hybrid IDS utilize a combination of existing signature-based systems and newly developed anomaly-based systems to protect against a broad range of attacks without the negative impact on performance. Despite their utility and wide applicability, Hybrid IDS have a number of limitations and challenges that remain unaddressed and require further attention. Only a handful of papers have been published to date focusing specifically on hybrid architectures that cover both detection mechanisms comprehensively [3]. Research addressing the specific challenges of Hybrid IDS themselves remains limited and needs to be expanded. For instance, research conducted on signature-based detection systems can directly influence anomaly-based detection systems, and vice versa. Very few IDS exploit the trade-off between the two classes of systems to develop better classifications that reduce the constraints in each. Moreover, few existing Hybrid IDS demonstrate sufficient adaptability to emerging attack methods.

The challenge of varying characteristics of different networks also needs further study [5]. It is important to develop adaptive hybrid systems that might adjust their models according to the characteristics of the monitored network. Such systems should be able to detect novel attacks without human intervention. Further studies are also needed to examine the impact an increase in anomalous traffic has on a Hybrid IDS. Comprehensive future research should be conducted on combining the two classes of systems into a single entity that autonomously handles both detection tasks. Data mining techniques have also been proposed as an effective way of discovering new attacks from audit trail data and of forming a new family of attacks that has a specific common behavior. These techniques should be investigated to determine their potential effectiveness as a new strategy to overcome the limitations of all current intrusion systems.

## 1. Limitations of Hybrid IDS

The analysis of the collected data brought to the detection requirements that could create the architecture for a hybrid intrusion detection environment. The proposed architectures have strengths and weaknesses with respect to different criteria. Even with a reduction in false positive alerts, IDS inevitably produce a certain level of acceptable false alarms. For that reason, the solution is the cooperation of IDS on the basis of false alerts analysis. The deployment of different technologies of IDS allows the matching of alerts that can complement each other and increase the precision of possible threats [9]. There would still be situations in which even a combination of different technologies would not be enough to avoid false alerts.

On the other hand, there are potential attacks that use specific mechanisms to explore the weaknesses of IDS, such as white-collar attacks. These attacks are performed by well-informed agents, with knowledge of the organization they are targeting and the techniques of the deployed security mechanisms, using this information to devise an intrusion that remains unnoticed by the IDS. Each IDS provides a partial view of the entire monitored environment. This is another area where the deployment of multiple IDS improves security. By complementing each other, they can provide a complete view of the environment. The observed strengths and weaknesses of a given architecture can determine the best possibilities for cooperation with other deployed architectures. In this sense, analysis of the possibility of communication and cooperation of the IDS deployed by the same organization is necessary [1].

## 2. Research Opportunities

This section highlights potential areas for future research and development within the domain of Hybrid IDS. The focus is specifically on addressing metrics such as the efficiency, performance, adaptability, and reduction of complexity; exploring the unexplored avenues of IDS which are the combinations of recent trending techniques such as ML and artificial intelligence or rules-based data filtering in Hybrid IDS secures world-wide applications; conducting comprehensive analysis and testing of different datasets on Hybrid IDS with comparative evaluation; and finally exploring the Hybrid IDS on Cloud platforms and OS for better services, which are currently in the burgeoning growth stage. Recent hybrid deep learning models developed for complex behavior analysis tasks demonstrate that combining temporal modeling, attention mechanisms, and ensemble learning can significantly enhance adaptability and detection accuracy, suggesting promising directions for next-generation Hybrid IDS research [29][30][31]. All of the above-stated unexplored opportunities could lead to the advancement of Hybrid IDS in terms of research and commercial viability, which is beneficial for both vendors and end users [5]. To ensure the robustness and future performance of Hybrid IDS systems, it is necessary to design possible scenarios for testing the efficiency and robustness of Hybrid IDS under various attacks, such as DDoS, Directory, and other malicious behaviors. Moreover, to avoid future loss of data, academics and industry researchers must analyze the vulnerability of rare attacks on the Hybrid IDS, which were not used for the testing of the performance of the Hybrid IDS modeled from the credibility of datasets [9].

Addressing these challenges and opportunities paves the way for further improvements and the broader applicability of Hybrid IDS.

## VII. CONCLUSION

The study and analysis of Hybrid IDS combining signature-based and anomaly-based detection as a second level of defense have led to several important findings, conclusions, and research perspectives. The analysis highlights that Hybrid IDS architectures integrating clustering techniques with supervised learning methods can achieve high detection performance when supported by effective preprocessing mechanisms. Cross-validation mechanisms are commonly employed in Hybrid IDS training processes to ensure robustness and generalization of detection models. Upon completion of the architecture and training of the models, Hybrid IDS are employed during the testing process, resulting in the detection of misuse and novel types of intrusions with a high degree of success, indicated by low misclassification rates and false-negative object counts.

The performance of the Hybrid IDS is further improved through the filtering of false alarms using the ensemble-learning approach. The counting of these false alarms is independently recorded for each of the models deployed in the ML-Hybrid IDS and concurrently fed back into the retraining process of the models. The retraining process is incremental and continuously runs along with the operational IDS and must be carefully managed not to incur any degradation in the overall IDS performance. The trained models are also evaluated in a continuous manner and must be able to account for any performance deterioration due to changes in the environment or the occurrence of new types of attacks. When performance degradation is observed, retraining strategies may be required to maintain detection effectiveness. This overall scheme is an overview of the ongoing research efforts. Future work includes, among others, the selection of a priori probability for each of the classes upon their first appearance within the environment of the IDS and the experimental evaluation of the effect of this probability on the run-time performance of the IDS, as well as the development of a new hybrid second level of detection using ensemble classifiers instead of a single decision tree for the anomaly detection level.

### 1. Summary of Findings

This subsection illustrates representative findings through a practical intrusion scenario frequently discussed in the Hybrid IDS literature. It offers a concise overview of the significant findings presented in the document.

A network hosts a web service that can be corrupted: that is to say, an intruder can initiate a series of operations/commands on the legitimate network merchant server such that the affected user accounts are tricked into approving unauthorized transactions, all while aiming to evade the scrutiny of the log analysis processes typically performed by the network IDS [1]. In an effort to thwart such activities, the network IDS models the legitimate behavior of the web service, the database, as well as the logging policies, and tasks the web service with a complementary logging mechanism able to record all executed database transactions [9]. Such logged operations can be modeled as a timeline of data record modifications: additions, deletions, as well as the alteration of data values. This sequence of modifications is assembled into a description of the event activity record stream. Detection of potentially misleading operations requires the definition of the unwarranted execution of events/commands, and also the grouping of event activity records comprising a relevant time window. That's where particular constraints about alterations of data record modification types are applied. By examining the database, user accounts are mutually linked; that is to say, the web service is able to disclose all transactions performed by a given user account. However, a small amount of legitimate wiring modifications can be performed on behalf of unrelated users. Each authenticated transaction is provided with a timestamp, and every legitimate grouping of modifications should abide by the temporal ordering of these timestamps.

### 2. Recommendations for Future Research

Hybrid IDS combines signature-based and anomaly-based detection mechanisms. Although this work makes an effort to come up with a Hybrid IDS that uses multi-technique detection methods, the use of only two datasets to assess the Hybrid IDS is a limitation. As there are many datasets available presently in the field of intrusion detection, it falls short of considering several well-known datasets in the assessment of its hybrid detection model. However, future work can consider some of the well-known and inadequate datasets like the UNSW-NB15 dataset and the ISCX-IDS2012 datasets on top of the two chosen datasets in this work. In consideration of this, it would be unrealistic to claim that the hybrid detection model is fully dataset-independent [9]. Since the score-based method is used to combine the output of multiple models, it, along with the models, can be varied for a suitable combination. Trained models can be saved and modified for future training, transfer learning, and a parcel of the best trained models can also be combined to provide a more robust detection model.

Another challenge that still remains unattended is a stateful analysis of the network traffic. This work conducted only a packet-based analysis of the sniffed packets in the network. Future work can consider inspecting the packets for a time window to detect attacks in a session-based environment. By doing this, denial of service attacks can be detected aside from sniffing the individual packets. Presently, many attacks like DDoS are being launched, and its detection is very hard with only the packet-based analysis. By maintaining the state of the sessions, the detection of such attacks can also be performed, which is a serious shortcoming in this work on top of the prospects already mentioned [1].

In conclusion, the proposed Hybrid IDS significantly enhances network security and provides a strong foundation for future research in intrusion detection.